\documentclass{PoS}

\usepackage{float}

\title{Parallel waveform extraction algorithms for the Cherenkov Telescope
    Array Real-Time Analysis}

\ShortTitle{Parallel waveform extraction algorithms for CTA}

\author{\speaker{Andrea Zoli},$^{a}$, Andrea Bulgarelli$^{a}$, Adriano De Rosa$^{a}$,
Alessio Aboudan$^{a}$, Valentina Fioretti$^{a}$, Giovanni De Cesare$^{a}$, Ramin Marx$^{b}$,
for the CTA Consortium\footnote{Full consortium author list at http://cta-observatory.org}\\
     \llap{$^a$}INAF/IASF Bologna, Bologna, Italy\\
     \llap{$^b$}Max-Planck-Institut f\"ur Kernphysik, Heidelberg, Germany\\
     E-mail:  \email{zoli@iasfbo.inaf.it}}
     
\abstract{The Cherenkov Telescope Array (CTA) is the next generation
observatory for the study of very high-energy gamma rays from about 20 GeV
up to 300 TeV. Thanks to the large effective area and field of view, the CTA
observatory will be characterized by an unprecedented sensitivity to transient
flaring gamma-ray phenomena compared to both current ground (e.g. MAGIC,
VERITAS, H.E.S.S.) and space (e.g. Fermi) gamma-ray telescopes. In order to
trigger the astrophysics community for follow-up observations, or being able
to quickly respond to external science alerts, a fast analysis pipeline is
crucial. This will be accomplished by means of a Real-Time Analysis (RTA)
pipeline, a fast and automated science alert trigger system, becoming a key
system of the CTA observatory. Among the CTA design key requirements to the
RTA system, the most challenging is the generation of alerts within 30 seconds
from the last acquired event, while obtaining a flux sensitivity not worse than
the one of the final analysis by more than a factor of 3. A dedicated software
and hardware architecture for the RTA pipeline must be designed and tested.
We present comparison of OpenCL solutions using different kind of devices like
CPUs, Graphical Processing Unit (GPU) and Field Programmable Array (FPGA) cards
for the Real-Time data reduction of the Cherenkov Telescope Array (CTA)
triggered data.}

\FullConference{The 34th International Cosmic Ray Conference,\\
		30 July-6 August, 2015\\
		The Hague, The Netherlands}

\begin{document}

\section{Introduction}
The Cherenkov Telescope Array (CTA) \cite{bib:cta} will be the biggest
ground-based very-high energy (VHE) $\gamma$-ray observatory, with a factor of
10 improvement in sensitivity compared to both current ground (e.g. MAGIC
\cite{bib:magic}, VERITAS \cite{bib:veritas}, HESS \cite{bib:hess}) and space
(e.g. Fermi \cite{bib:fermi}) gamma-ray telescopes considering in lower energy
range. To achieve this sensitivity three types of telescopes will be organized
in a grid. In order to cover the entire sky, the observatory will be divided
into two array of telescopes across two sites, one in each hemisphere.

Both arrays will be able to observe transient phenomena like Gamma-Ray
Bursts (GRBs) and gamma-ray flares. To capture these phenomena during their
evolution and for effective communication to the astrophysical community,
speed is crucial and requires a system with a reliable automated trigger that
can issue alerts immediately upon detection. This task will be performed by the
level-A analysis, also known as Real-Time Analysis (RTA), system of CTA
\cite{bib:bulgarelli} which is capable of triggering scientific alerts within
30 seconds from the beginning of an event. The RTA sensitivity will be not
worse than the final one by more than a factor of 3. The RTA alerts will be
used, also, to repoint part or the whole array to observe events that
cannot be otherwise possible.

RTA is a key component of the CTA on-site analysis infrastructure, and a huge
amount of computing power for the elaboration is forseen. This work benchmarks
both performances and performances per Watt, of different High Performance
Computing (HPC) solutions for the RTA waveform extraction, considering both
the temporal constraints and the current CTA prototyping data flow.

\section{Waveform extraction algorithm}

The waveform extraction algorithm is the first step of the RTA pipeline.
We used a six-sample sliding window method, one for each waveform, to find the
window with the maximum sum of samples and its related time \cite{bib:magicextract}.
The extracted $t$ time associated to the signal window is computed by the mean
of the sample time values, weighted by the sample value, using the following
equation:
\\

$t = \frac{\sum_{i=i_0}^{i_0+ws-1} s_i t_i }{\sum_{i=i_0}^{i_0+ws-1} s_i}$
\\

\noindent being $i$ the sample index within the selected window, $ws$ the window
size, $s_i$ the sample value and $t_i$ the $i$-sample signal time.

The algorithm input is the camera data generated from the CTA PROD2 simulations
\cite{bib:prod2} and saved using the Streaming Data Format (SDF) \cite{bib:sdf}.
In order to simulate correctly an RTA pipeline and access the camera raw data we
used the PacketLib C++ library, preloading the raw camera packets into a circular
buffer. Each raw camera packet contains 1141 or 1855 pixels, depending on the
camera type, with a fixed number of samples of 40 or 30 for each waveform respectively.
Considering the expected triggered camera event rate of 41 kHz, and a mean packet
size of 100 kB, samples of two bytes, the waveform extraction algorithm must
process 4.1 kB/s of events.

The sequential algorithm version extracts the waveforms per-event
using a window of $M$ and an input size of $N = WS$, being $W$ the number of
pixels and $S$ the number of samples of a triggered camera event. The final
complexity without optimizations it is $O(NM)$. We avoided the window loop,
and so the $M$ complexity, reusing the computed sums of the previous windows.
The final complexity of our best sequential algorithm is then $O(N)$.
We optimized also the case when $N$ is little, splitting the algorithm
into two parts: the first computes the sums and their relative times while the
second one finds the maximum sum using a reduction, with a lower complexity.

To run the algorithm on multiple threads we used OpenMP \cite{bib:openmp}.
The complexity of the parallel algorithm is $O(N / P)$ where
$P$ is the number of threads. In our scenario, the parallel access from multiple
threads to the same buffer, represent a critical section. Notice that in a
real case the performances and the scalability of the algorithm are in general
worse than the theorical case.

A GPU or an FPGA solution can perform better, so, we tested the algorithm using
OpenCL \cite{bib:opencl}. OpenCL is an open standard framework for
writing programs that execute across heterogeneous platforms, providing parallel
computing using task-based and data-based parallelism. We implemented two
different kernels (C99-like methods) parallelizing on the pixels and on the slices
respectively. To reduce the number of OpenCL calls and use correctly the
accelerators, we combined multiple events into groups. Testing different group
sizes we have found a good trade-off between the buffer size and minumum performances
required of 100 events per group regarding the GPU and 10k events for the FPGA.
This method emulates the real buffering of the data into different queues depending
on the camera type. For the benchmarks we counted the event grouping latency due
to the copy. We have performed some optimization, like the coalescent access to the
global device memory. For the FPGA we implemented a single work-item kernel
unrolling the windows loops, with a resulting board usage of 80\%.

Each test was run over the input circular buffer for 100K triggered camera events,
considering the entire event loop timing, including the parallel access to the
input buffer, the waveform extraction and the data transfer between cpu and
accelerators (only using OpenCL). The used test machine has the following tested
devices: an Intel Xeon E5-2697 v3 CPU with 132 GB of RAM DDR3, an Nvidia Tesla
K40 GPU and the Altera Stratix V GT FPGA. The machine runs CentOS 6.5 with gcc 4.8.2.
The OpenCL device drivers versions installed are OpenCL 1.2 (Build 57),
provided by the Intel OpenCL Runtime 15, OpenCL 1.1 CUDA 6.5.48 provided by
CUDA 6.5.14 and OpenCL 1.0 Altera SDK for OpenCL, Version 14.1.1 provided
by the Altera SDK for OpenCL \cite{bib:altera}. The sequential and the OpenMP
versions are optimized using -O2 compiler optimization options.

\newpage

\section{Results and analysis}

\begin{table}[H]
\centering
\begin{tabular}{|l|c|r|r|r|r|}
\hline
\textbf{Algorithm} & \textbf{Device} & \textbf{Nominal Power} & \textbf{Watt} & \textbf{Performance} & \textbf{Performance per watt} \\
 & & watt & watt & kevents/s (GB/s) & kevents/s/watt \\
\hline
Sequential      & CPU  & 147 &  47 &   6.66 ( 0.71) & 0.141 \\
\hline
OpenMP 8 cores  & CPU  & 147 & 137 &  46.80 ( 4.96) & 0.341 \\
OpenMP 56 cores & CPU  & 147 & 291 & 164.40 (17.43) & 0.861 \\
\hline
OpenCL          & CPU  & 147 & 248 &  25.43 ( 2.70) & 0.102 \\
                & GPU  & 163 &  95 &  36.97 ( 3.91) & 0.389 \\
                & FPGA & 164 &  21 &  10.93 ( 0.91) & 0.520 \\
\hline
\end{tabular}
\caption {The collected results for the sequential algorithm on a single CPU core
and for the parallel algorithms developed with OpenMP and OpenCL. The nominal power
is the power consumption of the test machine at nominal state. The watt column is
the power consumption difference from nominal state while running the tests. The target
performances are of 40 kevents/s. Higher performances per watt (power efficiency) are better.}
\label{table:bench}
\end{table}

The results of our tests are reported in Table \ref{table:bench}. We have
obtained optimal results in terms of performances using OpenMP using all the 56 cores
of the Intel CPU. Regarding the OpenCL solutions, more than half of the time is spent on data
transfer to devices, plus an additional time is spent on the host to group the events.
The OpenCL kernels execution is memory-bound, with most of the time spent loading
and storing data on the device memory. Considering these problems, we can
say that GPUs performed quite well, almost reaching the goal of 4 GB/s without much
optimization. On the contrary, the given FPGA OpenCL solution gives currently
poor performances. The reasons are mainly the kernel design and the bus
bottleneck. The current board design is based on a single work-item kernel,
exploiting the pipeline parallelism obtained from the window loop unrolling.
Better results can be obtained using the ``multiple compute units" optimization.
With this kernel and using the Altera channels for communication between
kernels we can theorically double the performances. Another way to greatly improve
the performances can be the use of a double buffer to parallelize the data
transfert to the device and the kernel execution. This improvement applies also to
the GPU case. Regarding the other problem of the bus access, instead, we have used
the Altera SDK for OpenCL with the Nallatech OpenCL board Intellectual Property (IP).
By now, it is not available a PCI-Express v3 bus IP, but only the PCI-Express v2 IP one.
So, even if the FPGA is physically attached to a PCI-Express v3 bus,
the memory bandwidth reaches only 2.5GB/s on both directions (tested using
the provided device memcopy test). This bottleneck should be overcomed with the
future IP releases. Using the PCI-Express v3 IP the data transfer should
be enough to reach the performance goal. It should be even better using the CAPI bus,
a a dedicated solution developed by IBM to increase the performances of the PCI-E buses
\footnote{\href{http://www-304.ibm.com/webapp/set2/sas/f/capi/home.html}
{http://www-304.ibm.com/webapp/set2/sas/f/capi/home.html}}.

In terms of performances per watt we reached the best results using the FPGA solution.
Given a single watt, the FPGA performs 34\% better then 8 CPU cores,
which are the required to sustain the target event rate of 41 kevents/s. The GPU
seems to be quite efficient too, but we its nominal consumption has to be subtracted,
so the real consumption is greater then the reported one.

\newpage

\section{Summary and conclusion}

The heterogenous computing is a key factor to maximize the performances of
the RTA pipeline. More importantly, each algorithm can be tailored to run over
different kind of devices, obtaining a solution that is also energy efficient.
This is important considering that the RTA pipeline will run on-site, but cutting
the maintenance costs could be useful even for the CTA software running off-site.

With this work we tested the waveform extraction algorithm using different kind
of device and methods. We reached optimal results with the CPU in terms
of performances, good ones with GPU and further tests for the FPGA are required.
Regarding the energy efficiency, instead, we reached the best results with the
FPGA, 34\% worse with the CPU and even worse with the GPU.

We plan to to test a better OpenCL kernel for the FPGA as soon as PCI-Express v3
bus IP will be released from Nallatech, we will test a double buffer solutions
for OpenCL, we will extend the tests using a IBM Power8 machine and use lower-budget
GPUs, we will evaluate the other RTA algorithms. The development of other algorithms
for the RTA are in progress. Considering that this problems are more cpu intensive,
we expect to see bigger advantages using GPU or FPGA devices.

\section{Acknowledgments}

We gratefully acknowledge support from the agencies and organizations listed under
Funding Agencies at this website: \href{http://www.cta-observatory.org/}{http://www.cta-observatory.org/}.

\end{document}